\documentclass[12pt,preprint2]{aastex}

\begin{document}

\title{New Elemental Abundances for V1974 Cygni}

\author{K.M. Vanlandingham}
\affil{Columbia University, Department of Astronomy, New York, NY 10027}
\email{kmv14@columbia.edu}

\author{G.J. Schwarz}
\affil{Steward Observatory, University of Arizona, Tucson, AZ 85721}
\email{gschwarz@as.arizona.edu}

\author{S.N. Shore}
\affil{Dipartimento di Fisica "Enrico Fermi", Universit\`a  di Pisa,
  largo Pontecorvo 3, Pisa 56127, Italy}
\email{shore@df.unipi.it}

\author{S. Starrfield} \affil{Department of Physics \& Astronomy,
Arizona State University, Tempe, AZ 85287-1504}
\email{sumner.starrfield@asu.edu}
 
\and

\author{R.M. Wagner} \affil{Steward Observatory, University of
Arizona, Tucson, AZ 85721} \email{rmw@as.arizona.edu}

\begin{abstract}
We present a new analysis of existing optical and ultraviolet spectra
of the ONeMg nova V1974 Cygni 1992.  Using these data and the
photoionization code Cloudy, we have determined the physical
parameters and elemental abundances for this nova.  Many of the
previous studies  of this nova have made use of incorrect analyses and
hence a new study was required.  Our results show that the ejecta are
enhanced, relative to solar, in helium, nitrogen, oxygen, neon,
magnesium and iron.  Carbon was found to be subsolar.  We find an
ejected mass of $\sim 2\times10^{-4}$M$_{\odot}$.  Our model results
fit well with observations taken at IR, radio, sub-millimeter and
X-ray wavelengths.

\end{abstract}


\section{Introduction}

Novae explosions are the result of a thermonuclear runaway (TNR)
occurring on the surface of a white dwarf (WD) in a close binary
system.  Material is transferred from the secondary star, a late-type
main-sequence star, through the inner Lagrangian point to an accretion disk and then
onto the WD.  Once enough material is accreted nuclear fusion begins
in the surface layers of the WD.  Since the WD is degenerate, this
leads to a TNR that then results in the ejection of the accreted
material.  Analysis of the ejecta provides information about the
physics of the nova process and the WDs on which they take place.

V1974 Cygni 1992 (hereafter Cyg 92) was discovered on 1992 February 19
by Collins (1992) (taken to be t$_0$ in this paper).  At that time it
was the brightest nova since V1500 Cygni and hence was one of the most
extensively observed novae in history with observations spanning the
entire spectral range from gamma-rays to radio.  Ultraviolet (UV)
observations made with the International Ultraviolet Explorer
satellite (IUE) show that Cyg 92 was a "neon" nova - one that takes
place on an ONeMg WD.

Three detailed abundance analyses have been carried out for Cyg 92:  Austin
et al. (1996, hereafter A96), Hayward et al. (1996) \& Moro-Mart\'{i}n et al. (2001).
Unfortunately these analyses were based on data that was dereddened incorrectly.  In the
initial analysis by A96 the reddening correction was applied in the
wrong direction for the optical spectra.  In other words, the optical spectra
were {\it reddened} rather than {\it dereddened}.  The UV spectra were
dereddened correctly, however since the analyses relied on the flux ratios
relative to H$\beta$ the UV ratios were ultimately affected by this mistake.
Hence a new analysis is required.
Using optical and UV data we have determined physical
parameters and elemental abundances for Cyg 92.  While we have used the same
photoionization code as A96 and Moro-Mart\'{i}n, we have developed a two-component
model to simulate the inhomogeniety of the nova shell which is more physically
accurate than these other analyses.  Hayward et al. also used a multi-component
model in their work however we have included significantly more data in our
analysis than they did.  We have compared our results
with X-ray, infrared (IR), sub-millimeter, and radio data in the literature and
find them consistent.

In Section 2 we briefly describe the observations used in our
analysis.  Section 3 reviews the reddening to the nova.  An overview
of our analysis technique is given in Section 4.  Section 5 contains a
detailed description of our model results and these results are
compared to others in the literature in Section 6.  Our conclusions
and summary are given in Section 7.

\section{Observations}

The UV data were obtained with IUE.  Low-dispersion large-aperture
data were taken with both  the Short Wavelength Primary (SWP:
1200-2000\AA) and the Long Wavelength  Primary (LWP: 2000-3400\AA)
cameras (resolution 7\AA).  The data were  reduced at the Goddard
Space Flight Center (GSFC) Regional Data Analysis  Facility (RDAF)
using the NEWSIPS IUE software.  The optical data were taken  with the
Perkins 1.8-m telescope at Lowell Observatory using the Ohio State
University Boller \& Chivens spectrograph (resolution 6\AA, wavelength
range  3200-8450\AA).  For a detailed description of the data and
spectral evolution see A96.  The data we are re-analyzing are the
spectra taken roughly 300, 400 and 500 days after the outburst.

The optical spectra were not absolutely flux calibrated.  In order to
combine them with the UV data we had to scale the optical flux to
match in the overlap region.  This was done using  the ratio of the
\ion{He}{2} lines at 1640\AA\ and 4686\AA.  Osterbrock (1989) gives
the theoretical ratio of these lines as 6.79.  After correcting the
spectra for reddening, we scaled the optical flux until the 1640/4686
ratio was equal to 6.79.  The dereddened line fluxes are given in
Table 1.

\begin{table}
\caption{Dereddened Observed Line Fluxes\tablenotemark{a}}
\vspace{1mm}
\begin{tabular}{@{}lrrr}
\hline
Line ID&Day 300&Day 400&Day 500\\
\hline
\ion{N}{5} 1240\tablenotemark{b}&1616.5&7203.98&2284.19\\
\ion{O}{4}$]$ 1402&526.25&117.49&26.92\\
\ion{N}{4}$]$ 1486&865.56&202.34&49.46\\
\ion{C}{4} 1549&309.94&63.06&18.17\\
$[$\ion{Ne}{5}$]$ 1575&147.08&34.67&10.30\\
$[$\ion{Ne}{4}$]$ 1602&403.26&96.16&21.71\\
\ion{He}{2} 1640&233.66&56.89& 18.29\\
\ion{O}{3}$]$ 1663&221.33&36.22&13.08\\
\ion{N}{3}$]$ 1750&379.42&64.90&17.98\\
\ion{C}{3}$]$ 1909&105.75&19.13&4.50\\
$[$\ion{Ne}{4}$]$ 2424&66.69&64.70& -\\
\ion{Mg}{2} 2798&221.02&59.73&5.9\\
$[$\ion{Ne}{5}$]$ 2976&100.11&57.96 &5.1\\
$[$\ion{Ne}{5}$]$ 3346\tablenotemark{c}&1627.95&612.36&372.28\\
$[$\ion{Ne}{5}$]$ 3426&4411.55&1833.24&1123.39\\
$[$\ion{Ne}{3}$]$ 3869&937.85&273.665&151.16\\
$[$\ion{Ne}{3}$]$ 3968&299.56&80.33&43.48\\
\ion{H}{1} 4102&62.64&13.25&6.44\\
\ion{H}{1} 4340&35.74&9.53&6.47\\
$[$\ion{O}{3}$]$ 4363&308.89&90.33&53.12\\
\ion{He}{2} 4686&53.91&18.59&15.80\\
$[$\ion{Ne}{4}$]$ 4721&199.00&62.34&31.63\\
\ion{H}{4} 4861&85.72&24.03&13.94\\
$[$\ion{O}{3}$]$ 4959&199.60&73.29&58.23\\
$[$\ion{O}{3}$]$ 5007&568.70&201.02&156.35\\
$[$\ion{N}{2}$]$ 5755&16.62&9.87&9.63\\
\ion{He}{1} 5876&10.06&1.64&1.17\\
$[$\ion{Fe}{7}$]$ 6087&16.72&8.13&9.94\\
$[$\ion{O}{1}$]$ 6300&4.58&1.53&1.96\\
\ion{H}{1} 6563&250.79&52.20&31.11\\
$[$\ion{O}{2}$]$ 7325&26.26&6.47&3.42\\
\hline
\tablenotetext{a}{Lines have been dereddened using E(B-V)=0.32.}
\tablenotetext{b}{UV fluxes are 10$^{-12}$ erg sec$^{-1}$ cm$^{-2}$}
\tablenotetext{c}{Optical fluxes are relative.  Scale factors for day 300, 400
and 500 are 6.4$\times10^{-13}$, 4.5$\times10^{-13}$, and 1.7$\times10^{-13}$, 
respectively.}
\end{tabular}
\end{table}

\section{Reddening}

A summary of reddening values used by other groups is given in Chochol
et al.  (1997, their Table 1).  Published values for E(B-V) range from
0.17 to 0.42.  For our analysis we have used two methods to determine
the reddening.  First we used the interstellar absorption feature at
$2175$\AA.  UV spectra taken with the Faint Object Spectrograph (FOS)
on the Hubble Space Telescope (HST) on 1995 November 30 clearly show
this feature.  We have taken these data and applied several different
reddening values using the extinction curve of Seaton (1979) and fit a
line through the continuum.  The spectra are shown in Figure 1.  These
data clearly favor higher reddening values with the best value being
E(B-V)=0.36.  For a second estimate of the reddening we looked at the
bolometric lightcurve.  If we apply a reddening correction of 0.36 we
find  that there is a deviation from constant flux.  If we instead
lower our reddening  value to 0.32 we are able to maintain a constant
flux while still removing most of  the interstellar absorption feature
in the UV.   Therefore we have chosen E(B-V)=0.32 as our reddening
value for our analysis.  If we apply this reddening value and the range
of distances to the nova from the literature (1.5-3 kpc) to the bolometric
flux (Shore et. al 1994) we find the luminosity of the nova to be 1-4$\times10^{38}$ erg
s$^{1}$ cm$^{-2}$.  This is roughly the Eddington luminosity for a 1M$_{\odot}$
WD.

\begin{figure}
\plotone{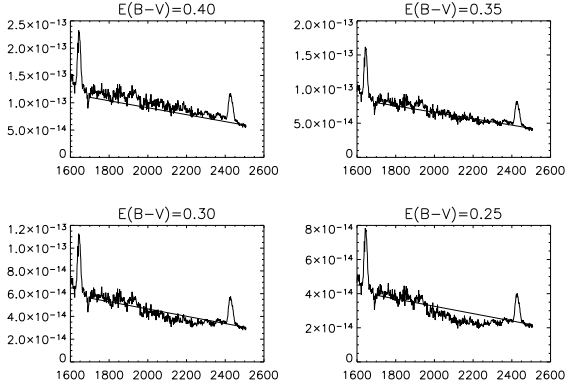}
\caption{The FOS spectra are shown here dereddened using several
different E(B-V) values.  A rough line is fit to the continuum  in
each plot.  These data clearly favor the higher reddening
values. \label{fig1}}
\end{figure}

The FOS spectrum also provided serendipitous support for our previous
study of the X-ray turnoff for V1974 Cyg (Figure 2). The FOS observation 
occurred soon after our last GHRS low
resolution spectrum (1995 Sept. 28) (Shore et al. 1997).  Comparing
this GHRS large aperture (2 arcsec) low resolution spectrum with the FOS
data shows that the emission line fluxes have changed significantly 
due to recombination following the X-ray turnoff, 
which occurred before the GHRS spectrum was obtained.   
At this stage in the evolution of the ejecta
($\sim$1300 days after outburst), the  helium recombination timescale
was approximately one month for the densities of about
1$\times10^6$cm$^{-3}$.  The integrated line flux of \ion{He}{2}
1640\AA\  decreases by about  20\% between these two observations that
are roughly two months apart,
from 1.26$\times 10^{-13}$ (GHRS) to 1.06$\times 10^{-13}$  erg
s$^{-1}$cm$^{-1}$ (FOS).  Since
both spectra were obtained  with the large aperture, which completely
contained the then-resolved ejecta, and the continua  agree in
intensity and slope, the change cannot be merely instrumental in
origin.  It thus  appears that the bulk of the emission came from
regions with densities that are characteristic of the clumps and
there was little emission from the diffuse gas (see also below).

\begin{figure}
\plotone{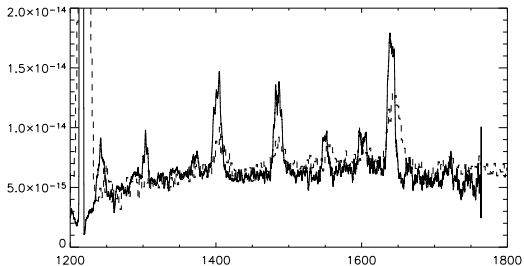}
\caption{A comparision of the FOS and GHRS spectra.  The GHRS spectrum
(solid line) was taken on 1995 Sept. 28 and the FOS spectrum (dashed
line) was taken on 1995 Nov. 30.  The flux in the  emission lines has
decreased significantly between the two observations.  No reddening
correction has been applied.}
\end{figure}

\section{Analysis}

We use Cloudy 94.00 (Ferland et al. 1998) to model the relative line
strengths of  Cyg 92 on three epochs independently.  In the past we
used this method to determine the  physical characteristics  of many
other novae (Vanlandingham et al. 1996, 1997, 1999; Schwarz et
al. 1997, 2001,  2002).  Cloudy 
simultaneously solves the equations of statistical  and
thermal equilibrium for specified initial physical conditions; the
model parameters are the spectral energy distribution of the
continuum  source, its temperature and luminosity, the hydrogen
density, the density law  for the ejecta (given by $\alpha$, where
$\rho \propto r^\alpha$), the inner and  outer radii of the shell, the
geometry of the shell, the covering and filling  factors of the shell,
the filling factor law (defined in the same way as the  density law)
and the elemental abundances (relative to solar).  We adopt a blackbody
to model the incident continuum.  Our previous work with other novae has
shown that using a model other than a blackbody for the underlying
continuum source results in little difference in the model fit for
this interval in the outburst (Schwarz 2002).  We ran models for one of
the dates using a NLTE model planetary nebula nuclei (Rauch 1997) as 
the continuum source instead of a blackbody and were able to achieve
the same fit to the data with changing only the temperature of the
source by less than 10\%.  While it has been found by Balman et. al (1998)
that blackbodies cannot be used to fit the soft X-ray observations,
we are not attempting to reproduce the spectral energy distribution of
the incident source so blackbodies give adequate results. 
We can constrain the radiation temperature of the source using published 
X-ray observations (Balman
et. al 1998), and the hydrogen density by observing the
relative strengths  of various ionization stages of a given element.
The FWHM of the emission lines and the terminal velocities of the P
Cygni  profiles provide the minimum and maximum velocities of the
ejecta and at any time since outburst, these are used to
determine an inner and outer radius of the nova  shell.  Based on the
luminosities derived for  other ONeMg novae, we choose a starting
value of 1$\times 10^{38}$ erg s$^{-1}$  and then allow the luminosity
to vary with successive iterations of the  code.  We assume a
spherical geometry for the shell, and start with a covering  factor of
unity.   We choose an initial value for the filling factor of 0.1,
since previous studies  have shown that novae do not eject homogeneous
shells but rather  clumps of gas imbedded in a diffuse gas (Shore et
al. 1993).

To determine the goodness of the fit of the model spectrum to the
observed  spectrum we use the $\chi^2$ of the model:
\begin{equation}
\chi^2 = \sum {\frac{(M_i-O_i)^2}{\sigma_i^2}}
\end{equation}
where $M_i$ is the modeled line flux ratio, $O_i$ is the observed line
flux  ratio, and $\sigma_i$ is the error in the measurement of the
observed flux for  each line (A96).  The error is determined by
measuring the line flux several times and looking at the variation of
the measurements.   The variation between measurements is primarily
due to the placement of the continuum and therefore the weakest lines
have the largest errors.  The flux for blended lines was estimated
using the 'deblend' option in the IRAF 'splot' package.  These lines
also have higher errors than the average.  These are typically on the
order of 20\% for the strongest lines but may be as high as 50\% for
the weakest or blended lines.  From our observations we typically have
$\sim$30 emission lines on which to base our fit.  Of the 24 input
parameters in  Cloudy, we fix 13: the density power law, the filling
factor and its power  law, the inner and outer radii, the geometry of
the shell,  and 7 abundances for which we had no data.  This usually
left us with $\sim$11 free  parameters and $\sim$19 degrees of
freedom.  A model is considered a good fit if it  has a reduced
$\chi^2$ (defined as the $\chi^2$ divided by the degrees of  freedom)
equal to one.

\section{Modeling the Spectra}

We modeled the same spectra described in A96 roughly corresponding to
300, 400 and 500 days after outburst.  The day 300 analysis is based
on the IUE spectra of 1992 December 4 combined with an optical
spectrum from 1992 December 15.  Day 400 is represented by the IUE
spectra from 1993 April 4 and an optical spectrum from 1993 March 16.
We encountered an apparent calibration error in the LWP spectrum for
this date.  In Figure 3 we have plotted the dereddened SWP and LWP
spectra for Day 400 with a line fit to the SWP continuum.  The LWP
continuum appears to be too low.  We can be somewhat confident that
the problem is with the LWP spectrum rather than the SWP spectrum
since there are two SWP spectra taken on this date and they agree with
one another.  If we multiply the LWP spectrum by a factor of ten then
the fit is much better.  While this is an eye estimate, the two
spectra match reasonably well if a multiplication factor anywhere
between 8 and 12 is used.  This uncertainty in the calibration of the
LWP spectrum results in a 20\% uncertainty in the line flux ratio for
the three emission lines obtained from this spectrum.  Lastly, our Day
500 analysis consists of the IUE spectrum from 1993 July 2 and an
optical spectrum from 1993 July 17.  There appeared to be a problem
with the LWP spectrum on July 2, most likely due to scattered light,
so only the SWP spectrum was used for this date.  The dates for the UV
and optical spectra do not match exactly for each of these pairings
however, due to the fact that the nova is evolving very slowly this
late after the outburst, this does not present a problem.

\begin{figure}
\plotone{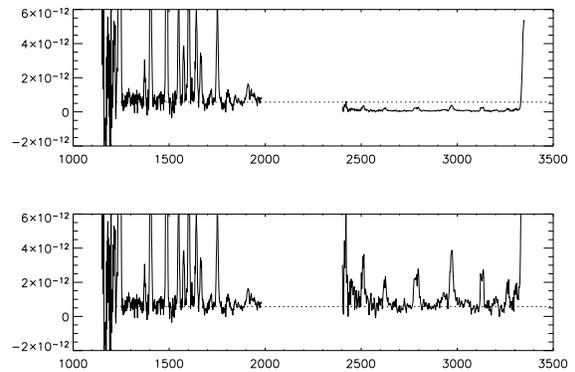}
\caption{The top box shows the SWP+LWP spectra for Day 400 with a
reddening correction of 0.32 applied.  A rough line has been fit to
the SWP continuum.  The bottom box shows the same spectra but with the
LWP flux multiplied by a factor of 10.  The fit to the LWP continuum
is much improved. \label{fig2}}
\end{figure}

\subsection{One-Component Model}

The first date we modeled was Day 300.
We used 29 emission lines in our fit (see Table 2).  The resulting values for
the physical parameters and elemental abundances are given in Table 4.
All parameters are in cgs unit.  The luminosity is given in erg s$^{-1}$, the density
in g cm$^{-3}$, and the radii are in cm.  All abundances are given by number relative to 
hydrogen relative to solar.  The fit has a
$\chi^2$=28 which gives a reduced $\chi^2$ of 1.6.  The number in parentheses next to
the abundance in Table 4 is the number of spectral lines that were used to determine that abundance.
The greater the number of lines the more constrained the value
of the abundance.  Thus the abundances of magnesium and iron are much more uncertain being
based solely on the fit to one spectral line, although they are probably accurate to within
a factor of 5-10.

\clearpage
\begin{table}
\caption{Day 300 Observed and Model Line Flux Ratios\tablenotemark{a}}
\vspace{1mm}
\begin{tabular}{@{}lrrrrr}
\hline
&Observed&1-component&&2-component&\\
Line ID&Ratio&Model Ratio&$\chi^2$&Model Ratio&$\chi^2$\\
\hline
\ion{N}{5} 1240&  29.33& 25.14&0.23&30.90&0.03\\
\ion{O}{4}$]$ 1402&  9.58&6.74&0.55&8.81&0.04\\
\ion{N}{4}$]$ 1486&  15.77&16.97&0.07&15.43&0.01\\
\ion{C}{4} 1549&   5.64&6.00&0.05&5.78&0.01\\
$[$\ion{Ne}{5}$]$ 1575&   2.68&2.08&0.55&2.50&0.05\\
$[$\ion{Ne}{4}$]$ 1602&   7.34&7.00&0.02&7.60&0.01\\
\ion{He}{2} 1640&   4.26&4.33&0.003&4.81&0.19\\
\ion{O}{3}$]$ 1663&   4.03&6.00&2.65&5.90&2.39\\
\ion{N}{3}$]$ 1750&   6.90&7.41&0.06&5.94&0.22\\
\ion{C}{3}$]$ 1909&   1.92&1.99&0.01&1.70&0.16\\
$[$\ion{Ne}{4}$]$ 2424&   1.22&0.21&2.73&1.43&0.13\\
\ion{Mg}{2} 2798&   4.02&3.85&0.02&3.99&0.00\\
$[$\ion{Ne}{5}$]$ 2976&   1.82&0.70&4.23&0.84&3.26\\
$[$\ion{Ne}{5}$]$ 3346&18.94& - & 10 &17.19&0.21\\
$[$\ion{Ne}{5}$]$ 3426&51.37& - & 10 &47.06&0.18\\
$[$\ion{Ne}{3}$]$ 3869&  10.93&14.40&2.52&11.32&0.03\\
$[$\ion{Ne}{3}$]$ 3968&   3.50&4.34&1.45&3.41&0.02\\
\ion{H}{1} 4102&   0.73&0.29&1.43&0.28&1.50\\
\ion{H}{1} 4340&   0.42&0.50&0.48&0.49&0.36\\
$[$\ion{O}{3}$]$ 4363&   3.61&2.58&0.91&2.69&0.71\\
\ion{He}{2} 4686&   0.63&0.56&0.14&0.64&0.01\\
$[$\ion{Ne}{4}$]$ 4721&   2.31&1.61&1.01&1.75&0.65\\
\ion{H}{4} 4861&   1.00&1.00&0.00&1.00&0.00\\
$[$\ion{O}{3}$]$ 4959&   2.33&0.67&3.18&0.83&2.60\\
$[$\ion{O}{3}$]$ 5007&   6.63&1.93&3.14&2.39&2.56\\
$[$\ion{N}{2}$]$ 5755&   0.19&0.16&0.27&0.14&0.98\\
\ion{He}{1} 5876&   0.12&0.12&0.002&0.07&0.90\\
$[$\ion{Fe}{7}$]$ 6087&   0.19&0.15&0.36&0.18&0.04\\
$[$\ion{O}{1}$]$ 6300&   0.05&0.06&0.18&0.09&3.65\\
\ion{H}{1} 6563&   2.93&3.26&0.13&2.99&0.00\\
$[$\ion{O}{2}$]$ 7325&   0.31&0.20&1.24&0.24&0.58\\
&&&&&\\
Total $\chi^2$&&&28\tablenotemark{b}&&22\\
\hline
\tablenotetext{a}{Ratios are relative to H$\beta$}
\tablenotetext{b}{$[$\ion{Ne}{5}$]$ 3346, 3426\AA\AA\ are not included
in the total $\chi^2$ of this model}
\end{tabular}
\end{table}

\clearpage

\begin{table}
\caption{Day 400 \& 500 Observed and Model Line Flux Ratios\tablenotemark{a}}
\vspace{1mm}
\begin{tabular}{@{}lrrrrrr}
\hline
&Observed&Day 400&&Observed&Day 500&\\
Line ID&Ratio&Model Ratio&$\chi^2$&Ratio&Model Ratio&$\chi^2$\\
\hline
\ion{N}{5} 1240&66.64  &84.80&0.83&96.37&92.58&0.02\\
\ion{O}{5} 1371&1.78&1.15&0.78&-&-&-\\
\ion{O}{4}$]$ 1402& 10.86 &9.99&0.04&11.36&4.29&2.42\\
\ion{N}{4}$]$ 1486& 18.71 &17.52&0.05&20.87&18.24&0.18\\
\ion{C}{4} 1549&  5.83 &6.67&0.23&7.67&7.57&0.00\\
$[$\ion{Ne}{5}$]$ 1575&  3.21 &2.71&0.26&4.34&1.72&4.04\\
$[$\ion{Ne}{4}$]$ 1602&  8.90 &4.19&3.11&9.16&2.80&5.36\\
\ion{He}{2} 1640&  5.26 &6.16&0.32&7.72&8.27&0.06\\
\ion{O}{3}$]$ 1663&  3.35 &4.39&1.07&5.52&2.68&2.94\\
\ion{N}{4} 1719&1.29&0.35&3.34&1.39&0.33&3.63\\
\ion{N}{3}$]$ 1750&  6.00 &4.12&1.09&7.58&5.51&0.83\\
\ion{C}{3}$]$ 1909&  1.77 &0.92&2.55&1.90&1.14&1.76\\
$[$\ion{Ne}{4}$]$ 2424&  5.98 &0.45&3.43&-&-&-\\
\ion{Mg}{2} 2798&  5.52 &3.28&1.03&-&-&-\\
$[$\ion{Ne}{5}$]$ 2976&  5.36 &0.91&7.67&-&-&-\\
$[$\ion{Ne}{5}$]$ 3346&25.50&21.49&0.62&26.70&16.48&3.66\\
$[$\ion{Ne}{5}$]$ 3426&76.32&58.83&1.31&80.58&45.13&4.84\\
$[$\ion{Ne}{3}$]$ 3869& 11.39 &11.58&0.01&10.84&10.18&0.09\\
$[$\ion{Ne}{3}$]$ 3968&  3.34 &3.49&0.05&3.12&3.07&0.01\\
\ion{H}{1} 4102&  0.55 &0.28&0.98&0.46&0.27&0.66\\
\ion{H}{1} 4340&  0.40 &0.49&0.58&0.46&0.48&0.02\\
$[$\ion{O}{3}$]$ 4363&  3.76 &3.96&0.03&3.81&2.53&1.24\\
\ion{He}{2} 4686&  0.77 &0.82&0.03&1.13&1.10&0.01\\
$[$\ion{Ne}{4}$]$ 4721&  2.60 &0.97&4.38&2.27&0.64&5.70\\
\ion{H}{4} 4861&   1.00&1.00&0.00&1.00&1.00&0.00\\
$[$\ion{O}{3}$]$ 4959&  3.05 &2.33&0.35&4.12&1.93&1.77\\
$[$\ion{O}{3}$]$ 5007&  8.37 &6.73&0.24&11.22&5.57&1.59\\
$[$\ion{N}{2}$]$ 5755&  0.41 &0.13&5.13&0.69&0.20&5.64\\
\ion{He}{1} 5876&  0.07 &0.08&0.18&0.08&0.07&0.27\\
$[$\ion{Fe}{7}$]$ 6087&  0.34 &0.40&0.21&0.71&0.88&0.32\\
$[$\ion{O}{1}$]$ 6300&  0.06 &0.11&3.10&0.14&0.13&0.02\\
\ion{H}{1} 6563&  2.17 &2.94&1.40&2.23&2.88&0.93\\
$[$\ion{O}{2}$]$ 7325&  0.27 &0.23&0.21&0.25&0.20&0.39\\
&&&&&\\
Total $\chi^2$&&&45&&&48\\
\hline
\tablenotetext{a}{Ratios are relative to H$\beta$}
\end{tabular}
\end{table}

\clearpage

\begin{table}
\caption{Nova Cyg 92 Model Parameters}
\vspace{1mm}
\begin{tabular}{@{}lrrrr}
\hline
&One-Component&Two-Component&Two-Component&Two-Component\\
Parameter&Day 300&Day 300&Day 400&Day 500\\
\hline
log(T$_{BB})$ K&5.52&5.52&5.67 &5.65\\
log(L) erg s$^{-1}$ &38.06&38.06&38.37 &38.1\\
log(H$_{den})$(diffuse) g cm$^{-3}$&-&7.097&6.99&6.79\\
log(H$_{den})$(clump) g cm$^{-3}$&7.75&7.75&7.39&7.1\\
$\alpha$&-3.0&-3.0&-3.0 &-3.0\\
log(R$_{in}$) cm &15.317&15.317&15.44 &15.54\\
log(R$_{out}$) cm &15.715&15.715&15.84 &15.94\\
Fill&0.1&0.1&0.1 &0.1\\
Power&0.0&0.0&0.0 &0.0\\
Cover(diffuse)&-&0.5&0.62&0.47\\
Cover(clump)&0.3&0.32&0.33&0.18\\
He\tablenotemark{a}&1.1(3)&1.0 (3)&1.3 (3) &1.4 (3)\\
C\tablenotemark{a}&0.40(2)&0.69 (2)&0.87 (2) &0.59 (2)\\
N\tablenotemark{a}&21.4(4)&35.5 (4)&57.6 (5)&41.7 (5)\\
O\tablenotemark{a}&3.9(7)&12.3 (7)&19.5 (8)&6.5 (7)\\
Ne\tablenotemark{a}&38.2(7)&56.5 (9)&44.9 (8)&23.0 (7)\\
Mg\tablenotemark{a}&1.3(1)&2.6 (1)&6.5 (1)&1.0 (0)\\
Fe\tablenotemark{a}&1.7(1)&1.3 (1)&3.8 (1)&9.5 (1)\\
$\chi^2$&28&22 &45 &48\\
&&&&\\
Total \# of lines &29& 31 & 33 & 29\\
\# of free parameters &11& 13 & 13 & 12\\
DOF &18& 18 & 20 & 17\\
Reduced $\chi^2$&1.6& 1.2 & 2.3 & 2.8\\
\hline
\tablenotetext{a}{Elements are given by number relative to hydrogen relative to solar.}
\end{tabular}
\end{table}

\clearpage

Despite the low $\chi^2$, there are several problems with this model.
First, the model is unable to adequately reproduce the high ionization
lines seen in the observations and in particular the $[$\ion{Ne}{5}$]$
3324, 3426\AA\ lines.  The $\chi^2$ quoted for the model is with these
two lines removed from the fit, each of which have an individual
$\chi^2$ of $\sim$10.  This is disconcerting since these are the
strongest lines in the optical spectrum and hence should have small
measurement errors associated with them.  There is some intrinsic
error introduced by the fact that these lines are at the bluest end of
the spectrum where the sensitivity of the CCD drops off.  However, the
magnitude of the discrepancy between the observed flux and the model
flux is too large to be explained by this.  There are also other
groups who report the flux ratios for these lines
(Moro-Mart\'{i}n et al. 2001) and our measurements agree with
theirs.  The difference between the model and observations is more
likely a shortcoming  of the model itself.  In addition to the
$[$\ion{Ne}{5}$]$ lines, our initial model is also unable to
reproduce the $[$\ion{Ne}{6}$]$ 76$\mu$/$[$\ion{Ne}{2}$]$ 128$\mu$
line ratio reported by Hayward et al. (1996).  Our model predicts a
value of 0.8 for this ratio while they measured a value of $\sim45$.
Clearly, the one-component model consistently under-predicts  the high
ionization lines in the spectra.

\subsection{A Two-Component Model}

Novae ejecta are not uniform in density but rather are clumpy, with
knots of high density material embedded in a more diffuse gas  (Shore
et al. 1993).  This is seen quite clearly in HST images of Cyg 92.
As a one-dimensional model,
Cloudy is not well suited to represent this type  of environment.  To
overcome this shortcoming, we have created a two-component model, one
component being the clumps and the other a  diffuse gas, where the
resulting line fluxes from the two components are then combined.
While this model is more realistic than a simple one-component model,
it is still not perfect since the two components are handled
separately by Cloudy when in reality they are not separate.  Ideally,
we would like to be able to embed the clumpy component within the
diffuse component but this is beyond the abilities of Cloudy.
However, we feel that our two-component  model is a reasonable
approximation until a better model is found.  Since our initial
one-component model fit a majority of the lines, we added a second
component to increase the flux of the high ionization lines.  Most of
the model parameters for the two components should be the same.  For
instance, the elemental abundances are not expected to vary between
the clumps and the diffuse component.  However, the physical
parameters, such as the density, filling factor and covering factor,
will necessarily be different for each component.  Going to a
two-component model increases the number of free parameters and makes
the task of finding a solution more difficult.  In our previous
analyses of other novae, one-component models fit the available
observations quite well.  This analysis of Cyg 92 has been the first
time a one-component model has had difficulty in fitting the
observations.  This is due to the wealth of data at wavelengths beyond
the optical and UV that we are able to use to constrain our models of
Cyg 92.

We added a diffuse component with a density that was less than the
original model (now considered the 'clump' component).  We again
adjusted the free parameters to obtain the best fit to the
observations.   The addition of the second component increased the
number of free parameters by 2  since we now have a second density and
a second covering factor.  The other parameters (shell radii,
elemental abundances, etc.) were set to the same value as the first
component.  The fit to the individual emission lines for our best
two-component model is shown in Table 3.  The parameters of the model
are  given in Table 4.  The fit has the same temperature and
luminosity for the underlying source as the  initial one-component
model.   The elemental abundances are slightly enhanced relative to
the one-component model.

The best values of the covering factors for the clump and  diffuse
components are found to be 0.32 and 0.5, respectively.  The reduced
$\chi^2$ of the  fit is 1.3 which is better than our original
one-component model and we now  include the $[$\ion{Ne}{5}$]$ lines in
the $\chi^2$.   The two-component model also improves our fit to the
IR lines ratios.  We find a ratio of
$[$\ion{Ne}{6}$]$/$[$\ion{Ne}{2}$]$=22 which is only a factor of 2 too
small instead of a factor of 50.  This shows that the two-component
model is significantly more realistic than  the simple one-component
model used previously.   Woodward et al. (1995) also found that
$[$\ion{Mg}{8}$]$ 30$\mu$/$[$\ion{Al}{6}$]$ 36$\mu$ $\sim$4 and our
two-component model predicts a ratio of 13.  The fact that this is now
higher than what is observed is most likely because we have set the
aluminum abundance to solar.  We have found in our work with other
ONeMg novae that aluminum is typically enhanced (Vanlandingham et
al. 1996, 1997, 1999).  We do observe lines of aluminum
($[$\ion{Al}{6}$]$ 2601\AA\ and \ion{Al}{2}$]$ 2665\AA) in the Cyg 92
spectra however they are too weak to measure reliably so we have not
used them in our models.  If we increase the abundance of aluminum to
2.4 times solar  then this ratio matches the observations.

As an additional check of our models, we can use the radio and
sub-millimeter  observations of Cyg 92.  The observations pertinent to
this analysis range from 1 to 500 GHz and were obtained within $\pm$
50 days of the Day 300 optical and UV observations (Hjellming 1996;
Ivison et al. 1993; Eyres, Davis, \& Bode 1996).  To  compare our
model to the observations, the luminosity was scaled to  distances of
1.5 and 3 kpc corresponding to the range of distances reported in the
literature.  Figure 4 shows the comparison of the model to the
observations.  The larger distance (solid line)  is consistent with
the average of the sub-millimeter observations on days 234 and 356 but
not the radio data.  At 1.5 kpc the model is in better agreement with
the radio observations but overestimates the sub-millimeter data.  The
disagreement may be due to a number of factors in the model including
being optically thin or having the wrong temperature.  In addition,
the radio images clearly showed an ellipsoidal shell whereas our
models are spherical.

\begin{figure}
\plotone{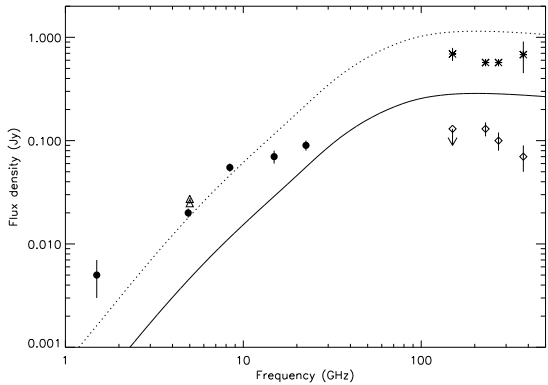}
\caption{Model for Day 300 as compared to the observed radio and
sub-millimeter data in the literature.  The radio data are taken from
Hjellming (1996) [filled circles] and Eyres et al. (1996) [day 270 and
315 = triangles].  The sub-millimeter data are from Ivison et
al. (1993) [day 234 = asterisks and day 356 = diamonds].  The dotted 
line represents the model luminosity scaled to a distance
of 1.5 kpc while the solid line is scaled to 3 kpc.
\label{fig3}}
\end{figure}

After finding a fit to the Day 300 data we then proceeded to use our
two-component model to fit days 400 and 500.  The individual emission
line fits for these days are shown in Table 3 and the model parameters
are given in Table 4.  These later dates are not fit as well as Day
300 as there is not as much data at other wavelengths to constrain the
fits.  There are some lines in Table 3 that have much large $\chi^2$
than most of the others.  Some of these, such as $[$\ion{Ne}{4}$]$]
1602\AA\ and $[$\ion{Ne}{4}$]$ 4721\AA, are probably a result of the
line being  blended making an accurate measurement of the flux
difficult.  Other lines, such as \ion{N}{4} 1719\AA\ and
$[$\ion{N}{2}$]$ 5755\AA, are weak lines which, again, increases the
error  in the measurement of the flux.  We may have underestimated the
errors for these lines in our calculation of the $\chi^2$.

If we compare the results from all three dates, we notice some trends.
The effective temperature of the continuum source increases slightly
with time.  This is as expected as the ejected shell expands and
reveals more of the underlying WD surface.  The luminosity is roughly
constant, as is the covering factors of the two components.   This is
in agreement with the findings of Balman et al. (1998).  Their X-ray
observations show the effective temperature peaking at day 511 and a
constant bolometric luminosity from day 255-511.  The nova then turned
off shortly  after this at roughly day 550.  The abundances values for
all three dates typically agree within a  factor of 2.5.  All the
abundances except for carbon are enhanced relative to solar values.

Based on the parameters determined from our three dates we can
calculate the hydrogen ejected mass predicted by our two-component
models.  In order estimate the ejected mass we take the shell defined
by the inner and outer radii and divide it into 1000 nested shells.
The density of the innermost shell is set to the starting density of
the model and is then progressed based on the density law of the
model.  The filling factor is applied in the same manner.  The
resulting mass is then multiplied by the covering factor.  Using this
method, we find an ejected mass of $2.1\pm0.2\times10^{-4}M_{\odot}$,
$2.8\pm0.3\times10^{-4}M_{\odot}$, and $2.1\pm0.2\times10^{-4}M_{\odot}$ for days
300, 400 and 500, respectively.  The ejected mass is roughly constant
over all dates, which is as expected.  Shore et al. (1993) finds that
the ejected mass can be calculated as $10^{-4} Y^{-1/2} M_{\odot}$,
where Y is the average enhancement factor for the helium abundance.
If we use this equation and an average helium abundance from our three 
models, we find $M_{ej}=1.9\times10^{-4}M_{\odot}$, which agrees with
our model values.  Our masses agree fairly well with
this calculation.  Other groups have found ejected mass estimates for
Cyg 92 (Shore et al. 1993, Krautter et al. 1996, Woodward et al. 1997)
in the same range that we find here.

\clearpage
\begin{table}
\caption{Abundance Comparison} 
\vspace{1mm}
\begin{tabular}{@{}lccccc}
\hline
& Average from&&&\\
Parameter&this paper \tablenotemark{a}&A96 & Moro-Mart\'{i}n& Hayward&Paresce\\
\hline
He&1.2$\pm$0.2&4.4 &4.5&4.5 \tablenotemark{b} &$\sim$2\\
C&0.7$\pm$0.2& -  & 70.6 \tablenotemark{b}&12 \tablenotemark{b}&\\
N&44.9$\pm$11&282 &50.0&50 \tablenotemark{b}&\\
O&12.8$\pm$7&110 &80.0&25 \tablenotemark{b}&2-4\\
Ne&41.5$\pm$17&250 &250.0&50&15-27\\
Mg&4.6$\pm$3&- &129.4 \tablenotemark{b}&5&\\
Al&$>$ 1.0\tablenotemark{c}&- &127.5 \tablenotemark{b}&5 \tablenotemark{b}&\\
Si&-&- &146.6 \tablenotemark{b}&6 \tablenotemark{b}&\\
S&-&- &1.0&5 \tablenotemark{b}&\\
Fe&4.9$\pm$4&16&8.0&4\tablenotemark{b}&\\
\tablenotetext{a}{Errors given are 1$\sigma$}
\tablenotetext{b}{No lines of this element are present in their spectra}
\tablenotetext{c}{This is a rough estimate based on the IR line ratios}
\end{tabular}
\end{table}
\clearpage

\section{Comparison to other results}

The first extensive analysis of the optical and UV data was done by
A96 using an older version of the Cloudy code.  As mentioned earlier, 
A96 made an error in applying the reddening corrections to their fluxes.  
Moro-Mart\'{i}n et al. (2001) and Hayward et al. (1996) also
conducted abundance analyses of Cyg 92 using Cloudy.  Unfortunately,
both of these groups based their modeling on the results of A96 and
hence propagated the errors from that analysis into their  work.  In
addition,  both of these studies report abundance values for elements
which are not  represented in their spectra.  It is possible to
predict an upper limit  for a given element by increasing its
abundance until the model produces emission  lines that should have
been seen but are not seen in the data.  These two groups, however,
report such large abundances  of these unseen elements that emission
lines would have been easily seen in the spectrum.

A third analysis is found in the literature.  Paresce et al. (1995)
gave rough abundances values for a few elements determined by using
Cloudy on a specific "knot" of material from their HST spectra.  These
results, shown in Table 5, did not rely on the analysis of A96.  They
state that their results are lower limits on the abundances.

Finally we note that Shore et. al (1997) took the results from A96 and propagated them
forward in time to Day 1300.  Using GHRS data, with a much higher S/N
than that of A96, they noted that the carbon abundance found by A96 was
too high.  

Given the error in A96 a comparison between our results and those of
Hayward and Moro-Mart\'{i}n would be misleading.  It is not
surprising that  our abundance results do not agree with their
results.  The discovery of the error in A96 was one  of the primary
motivations for our re-analysis of Cyg 92.  Table 5 shows our results
along with those of A96, Hayward and Moro-Mart\'{i}n.  In
general, our abundances are much lower than those of the three groups.
Our results are higher than those found by Paresce, however, they are
not in disagreement since Paresce's numbers are lower limits.

Through our previous work we have found striking similarities between many 
of the ONeMg novae (Shore et. al 2003, Vanlandingham et. al 1999).  A complete
and thorough comparison between the abundances found here and other ONeMg
novae will be the subject of a future work.

\section{Conclusions}

We have applied a two-component photoionization code to three separate
observations of Cyg 92 and have derived the physical characteristics
of the ejecta on these dates.  Our initial one-component model was
unable to reproduce the high ionization  lines seen in the spectra.
By adding a second, low density, component to our models we were able
to correct this problem.  We find the ejecta to be enhanced, relative
to solar, in He, N, O, Ne, Mg and Fe.  Carbon is found to be subsolar.
Our models predict an ejected mass of $\sim 2\times10^{-4}M_{\odot}$
which is in agreement with  what has been found for other ONeMg novae.

Our results replace the earlier analysis of A96 that contained an
error in the reddening correction.  The two other analyses in the
literature (Hayward et al. (1996) \& Moro-Mart\'{i}n et
al. (2001)) based their work on the results of A96 and so propagated
this error into their work.  Because there is such a large parameter
space and many of the parameters are interdependent it is difficult to
determine if a solution to one set of spectra is unique.  By modeling
three sets of observations independently, taken at different times
during the evolution of the nova shell, we increase the confidence in
our solution.  If all three days arrive at the same abundance
solution, then we can have much more confidence that it is the true
solution.  The fact that our models are able to fit the IR, radio,
sub-millimeter and X-ray observations further strengthens our
conclusions.

\acknowledgments

The authors would like to thank G. Ferland for use of his Cloudy
photoionization code and J. Aufdenberg and J. Jos\`e 
for many useful discussions.  We would also like to thank the referee
for a very thorough review.  The referee's comments have greatly improved 
the quality of the paper.
S. Starrfied acknowledges partial support from NSF and NASA grants to
ASU;  S. N. Shore acknowledges partial support from NASA grants to
IUSB, and INAF 2002 and INFN/Pisa.

\end{document}